\documentclass[11pt,a4paper]{article}%
\usepackage{amsfonts}
\usepackage{amsmath}
\usepackage{amssymb}
\usepackage{graphicx}
\usepackage{geometry}
\usepackage{hyperref}
\hypersetup{
    colorlinks=true,
    linkcolor=blue,
    filecolor=magenta,      
    urlcolor=cyan,
}

\providecommand{\U}[1]{\protect\rule{.1in}{.1in}}

\makeatletter
\def \@removefromreset#1#2{\let \@tempb \@elt
\def \@tempa#1{@&#1}\expandafter \let \csname @*#1*\endcsname \@tempa
\def \@elt##1{\expandafter \ifx \csname @*##1*\endcsname \@tempa \else
\noexpand \@elt{##1}\fi}     \expandafter \edef \csname cl@#2\endcsname{\csname cl@#2\endcsname}     \let \@elt \@tempb
\expandafter \let \csname @*#1*\endcsname \@undefined}

\@removefromreset{equation}{section}

\@removefromreset{theorem}{section}
\makeatother
\begin{document}

\title{Two-sequential Conclusive Discrimination between Binary Coherent States via Indirect Measurements}
\author{Min Namkung$^{1}$ and Elena R. Loubenets$^{1,2}$\\$^{1}$National Research University Higher School of Economics, \\Moscow 101000, Russia\\$^{2}$Steklov Mathematical Institute of Russian Academy of Sciences, \\Moscow 119991, Russia}
\maketitle

\begin{abstract}
A general scenario for an $N$-sequential conclusive state discrimination introduced recently in Loubenets and Namkung [arXiv:2102.04747] can provide a multipartite quantum communication realizable in the presence of a noise. In the present article, we propose a new experimental scheme for the implementation of a sequential conclusive discrimination between binary coherent states via indirect measurements within the Jaynes-Cummings interaction model. We find that if the mean photon number is less than 1.6, then, for our two-sequential state discrimination scheme, the optimal success probability is larger than the one presented in Fields, Varga, and Bergou [2020, IEEE Int. Conf. Quant. Eng. Comp.]. We also show that, if the mean photon number is almost equal to 1.2, then the optimal success probability nearly approaches the Helstrom bound.
\end{abstract}

\section{Introduction}
Since a coherent state is robust under an external noise and is easily experimentally implemented, it has been widely used as an information carrier for a quantum communication protocol \cite{g.cariolaro}. The main purpose of a quantum communication is to optimize the success probability for discriminating between several  states. Until now, a lot of experimental schemes for the optimal coherent state discrimination have been theoretically presented \cite{r.s.kennedy,s.j.dolinar,m.sasaki,r.s.bondurant,s.izumi,f.e.becerra,r.han,m.namkung}. Especially, the quantum communication protocols assisted by the probabilistic amplification have been proposed in \cite{m.rosati,h.adnane}. The influence on the performance of a quantum communication protocol of the phase diffusion \cite{j.trapani} quantum noise has been analyzed in \cite{h.adnane}.

Beyond a standard quantum state discrimination between a sender and a receiver, the sequential unambiguous state discrimination scenario between a sender and $N$ receivers was presented \cite{j.a.bergou} in 2013, and experimental schemes for implementation of this scenario of a state discrimination have been theoretically proposed \cite{m.namkung2,m.namkung3}. For example, when a sender prepares one of two polarized single photon states, then $N$ receivers can build their quantum measurements by using the Sagnac-like interferometers \cite{f.a.torres-ruiz,m.a.solis-prosser}. Also, when a sender prepares one of binary coherent states, then $N$ receivers can build their quantum measurements on the basis of the idea of Banaszek and Huttner in \cite{k.banaszek,b.huttner}. 

Unfortunately, an external noise may transform a coherent pure state to a mixed state, so that the sequential unambiguous state discrimination protocol in \cite{j.a.bergou} can be implemented only in an ideal case. Meanwhile, a sequential conclusive state discrimination, where every receiver's measurement outcome is always conclusive, can be implemented even in presense of noise. This means that a sequential conclusive state discrimination can provide a multipartite quantum communication realizable in a real world. In \cite{d.fields}, a sequential conclusive discrimination of two pure states was considered. 

Recently, a general framework for the $N$-sequential conclusive state discrimination has been presented in \cite{e.r.loubenets2}, which can be applied both for discrimination of pure or mixed quantum states and also for any number $N$ receivers. For this new scenario of a sequential conclusive state discrimination, experimental schemes should be theoretically developed. 

In the present article, we propose an experimental scheme for implementing sequential conclusive discrimination of binary coherent states via indirect measurements within the Jaynes-Cummings interaction model. We find that if the mean photon number is less than 1.6, then, for our two-sequential state discrimination scheme, the optimal success probability is larger than the one presented in \cite{d.fields}. We also show that, if the mean photon number is almost equal to 1.2, then the optimal success probability nearly approaches the Helstrom bound. This implies that our experimental scheme can provide a multipartite quantum communication protocol which succeeds with a high probability. Furthremore, we emphasize that our scheme for discrimination between binary coherent states is also successful in the presence of a noise.\footnote{The detailed analytical and numerical work analyzing the effect of the noisy environment is now in preparation.}

The results of the present article differ from \cite{m.namkung_coherent} in the following features. The indirect measurement considered in \cite{m.namkung_coherent} outputs an inconclusive outcome with a nonzero probability, meanwhile, that considered in the present article does not. Also, the indirect measurement in \cite{m.namkung_coherent} is constructed via linear optical devices, meanwhile, that considered in the present article is constructed via interaction between the light and the two-level atom.

The present article is organized as follows. In Section 2, we shortly recall the basics of a general framework for $N$-sequential quantum state discrimination which we have introduced in \cite{e.r.loubenets2}. In Section 3, we specify a general scenario for an $N$-sequential conclusive state discrimination which we have introduced in \cite{e.r.loubenets2} for the case of two receivers and receivers' indirect measurement described in the frame of the Jaynes-Cummings interaction model. In Section 4, we derive the expression for the success probability of the two-sequential conclusive discrimination between two coherent states via indirect measurements within the Jaynes-Cummings model and numerically investigate the optimal case. In Section 5, we summarize the main results.

\section{Preliminaries: Framework of $N$-sequential conclusive state discrimination}
In this section, we briefly introduce the preliminaries regarding the framework of $N$-sequential conclusive state discrimination (For detail, see the Section 2 and 3 in \cite{e.r.loubenets2}).

According to the mathematical framework introduced by \cite{e.b.davies,p.busch,a.s.holevo}, a quantum measurement is described by a state instrument $\mathcal{M}=\{\mathcal{M}(\omega),\omega\in\Omega\}$, where the values $\mathcal{M}(\omega):\mathcal{T}(\mathcal{H})\rightarrow\mathcal{T}(\mathcal{H})$ are completely positive bounded linear maps satisfying the following relation:
\begin{equation}
\sum_{\omega\in\Omega}\mathrm{tr}\left\{\mathcal{M}(\omega)\left[T\right]\right\}=\mathrm{tr}\left\{\mathcal{M}(\Omega)\left[T\right]\right\}=\mathrm{tr}\left\{T\right\}, \ \ T\in\mathcal{T}(\mathcal{H}).
\end{equation} 
Here, $\omega\in\Omega$ is the outcome observed by the quantum measurement. For the given quantum state described by a density operator $\rho$ on $\mathcal{H}$, the probability to observe the outcome $\omega$ by the quantum measurement is given by
\begin{equation}
\mu(\omega|\rho)=\mathrm{tr}\left\{\mathcal{M}(\omega)\left[\rho\right]\right\}.\label{new_1}
\end{equation}

$N$-sequential state discrimination is performed in a sequence $A|\rightarrow1\rightarrow...\rightarrow N$ including a sender Alice and $N$ receivers. Assume that Alice prepares a quantum state $\rho_i$ ($i=1,...,r$) with a prior probability $q_i$. Then, Alice sends the initial state
\begin{equation}
\rho_{in}=\sum_{j=1,...,r}q_j\rho_j \label{new_2}
\end{equation}
to the first receiver. Based on the above mathematical description, $N$ receivers constitute the consecutive quantum measurement described by the state instrument with an outcome $\omega=(j_1,...,j_N)\in\{1,...,r\}^N$:
\begin{equation}
\mathcal{M}_{A|\rightarrow1\rightarrow...\rightarrow N}(j_1,...,j_N)[\cdot]:=\mathcal{M}_N(j_N)[\mathcal{M}_{N-1}(j_{N-1})[...\mathcal{M}_1(j_1)[\cdot]...]].
\end{equation}
For a given initial state (\ref{new_2}), the probability to observe the outcome $(j_1,...,j_N)$ by the consecutive quantum measurement is given by the following expression in terms of (\ref{new_1}):
\begin{align}
\mu(j_1,...,j_N|\rho_{in})&=\sum_{j=1,...,r}q_j\mathrm{tr}\left\{\mathcal{M}_{A|\rightarrow1\rightarrow...\rightarrow N}[\rho_j]\right\}\nonumber\\
&=q_j\mathrm{tr}\left\{\mathcal{M}_N(j_N)[\mathcal{M}_{N-1}(j_{N-1})[...\mathcal{M}_1(j_1)[\rho_j]...]]\right\}.
\end{align}
Therefore, for the given states $\rho_1,...,\rho_r$ and the prior probabilities $q_1,...,q_r$, the probability that every $N$ receiver obtains the correct outcome takes the form
\begin{align}
\mathrm{P}_{A|\rightarrow1\rightarrow...\rightarrow N}^{success}(\rho_1,...,\rho_r|q_1,...,q_r)=\sum_{j=1,...,r}q_j\mathrm{tr}\left\{\mathcal{M}_N(j_N)[\mathcal{M}_{N-1}(j_{N-1})[...\mathcal{M}_1(j_1)[\rho_j]...]]\right\},\label{1}
\end{align}
which is so called \textit{success probability} of the $N$-sequential conclusive state discrimination.

\section{Two-sequential conclusive discrimination via indirect measurements}
In this section, we specify for the case of two receivers a general scenario for an $N$-sequential conclusive state discrimination which we have introduced in \cite{e.r.loubenets2}.

\begin{figure}[ptb]
\includegraphics[width=\linewidth]{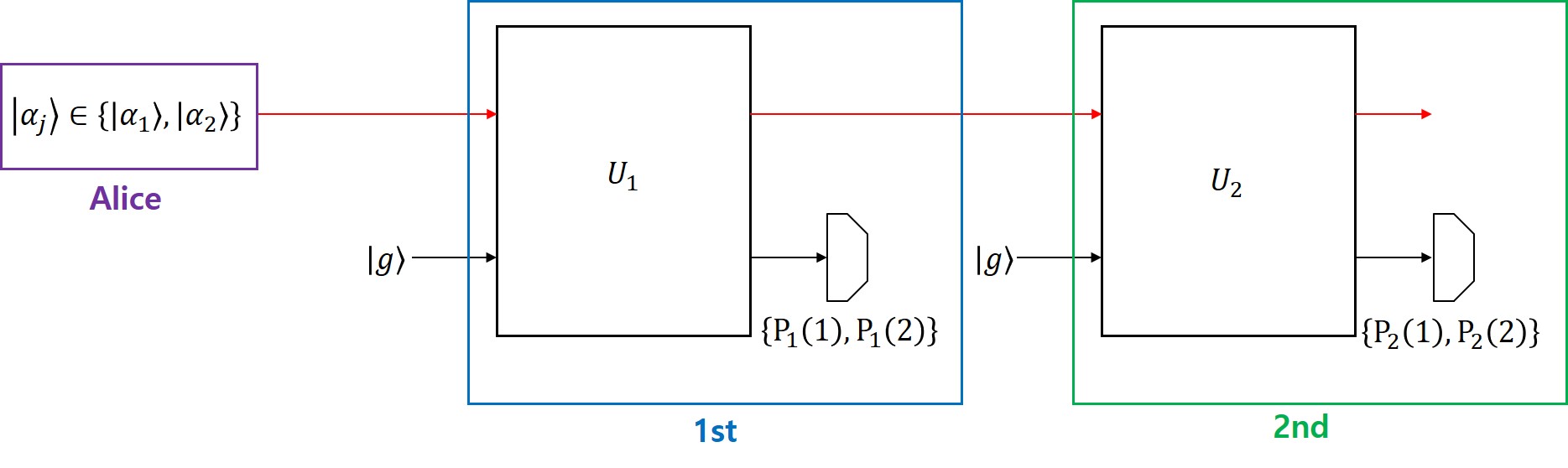}\caption{Description of two-sequential conclusive discrimination between binary coherent states. Here, the second receiver performs the indirect measurements on the posterior state conditioned by the first receiver's measurement outcome}
\end{figure}

Let Alice prepare one of two quantum states ${\rho}_1,{\rho}_2$ with prior probabilities $q_1,q_2$, and let $\mathcal{M}_l$ ($l=1,2$) be a state instrument \cite{a.s.holevo} describing a conclusive quantum measurement with outcomes $j\in\left\{1,2\right\}$ of each $l$-th sequential receiver. Then the consecutive measurement by two receivers is described by the state instrument \cite{e.r.loubenets2}:
\begin{equation}
\mathcal{M}_{A|\rightarrow1\rightarrow2}(j_1,j_2)[\cdot]:=\mathcal{M}_2(j_2)\left[\mathcal{M}_1(j_1)\left[\cdot\right]\right] \ \ j_1,j_2\in\{1,2\}, \label{0.5}
\end{equation}
and the success probability has the form 
\begin{align}
\mathrm{P}_{A|\rightarrow1\rightarrow2}^{success}(\rho_1,\rho_2|q_1,q_2)=\sum_{j=1,2}q_j\mathrm{tr}\left\{\mathcal{M}_2(j)\left[\mathcal{M}_1(j)\left[\rho_j\right]\right]\right\}.\label{1}
\end{align}
For details, see Eq.(27) in \cite{e.r.loubenets2}. 

Recall that according to the Stinespring-Kraus representation 
\begin{equation}
\mathcal{M}_l(j)[\cdot]=\sum_mK_{l}^{(m)}(j)(\cdot)K_{l}^{(m)\dagger}(j), \ \ \sum_{j,m}K_{l}^{(m)\dagger}(j)K_{l}^{(m)}(j)=\mathbb{I}_{\mathcal{H}},\label{1.5}
\end{equation}
where $K_{l}^{(m)}(j)$ are the Kraus operators for each $l$-th indirect measurement and in general, $m\in\{1,\cdots,m_0\}$. If $m_0=1$, then a state instrument is called pure and admits the representation
\begin{equation}
\mathcal{M}_l(j)[\cdot]=K_{l}(j)(\cdot)K_{l}^\dagger(j), \ \ \sum_{j}K_{l}^\dagger(j)K_{l}(j)=\mathbb{I}_{\mathcal{H}}.\label{3.33}
\end{equation}

Substituting (\ref{3.33}) to (\ref{1}), we have:
\begin{equation}
\mathrm{P}_{A|\rightarrow1\rightarrow2}^{success}(\rho_1,\rho_2|q_1,q_2)=\sum_{j=1,2}q_j\mathrm{tr}\left\{K_2(j)K_1(j)\rho_jK_1^\dagger(j) K_2^\dagger(j)\right\}.\label{17.7}
\end{equation}
If $\rho_i$ are pure states $\rho_i=|\psi_i\rangle\langle\psi_i|$, then
\begin{equation}
\mathrm{P}_{A|\rightarrow1\rightarrow2}^{success}(|\psi_1\rangle,|\psi_2\rangle|q_1,q_2)=\sum_{j=1,2}q_j\left|\left|K_2(j)K_1(j)|\psi_j\rangle\right|\right|_{\mathcal{H}}^2.\label{21.1}
\end{equation}

In our protocol, we realize the conclusive quantum measurement of each receiver via the indirect measurement described by the statistical realization\footnote{On the notion of a statistical realization, see, for example, in \cite{e.r.loubenets2}.}
\begin{equation}
\Xi_l:=\left\{\widetilde{\mathcal{H}},{\sigma}_l,\mathrm{ P}_l,{ U}_l\right\} \ \ l\in\{1,2\},\label{1.75}
\end{equation}
where $\widetilde{\mathcal{H}}$ is a two dimensional complex Hilbert space, $\sigma_l=|b_l\rangle\langle b_l|$ is a pure state on $\widetilde{\mathcal{H}}$, $\mathrm{ P}_l$ is a projection-valued measure $\{P_{l}(1),P_{l}(2)\}$ with values, 
\begin{equation}
\mathrm{ P}_l(j)=|\pi_l(j)\rangle\langle\pi_l(j)| \ \ j\in\{1,2\}, \ |\pi_l(j)\rangle\in\widetilde{\mathcal{H}}, \label{P}
\end{equation}
and a unitary operator \cite{e.r.loubenets2,e.r.loubenets3}
\begin{align}
U_l(|\psi\rangle\otimes|b_l\rangle)&=\sum_{j=1,2} K_l(j)|\psi\rangle\otimes|\pi_l(j)\rangle, \ \ \textrm{for each } |\psi\rangle\in\mathcal{H},\label{U}
\end{align}
for each $l=1,2$. Here,
\begin{align}
(\langle \phi|\otimes\langle\pi_l(j)|)U_l(|\psi\rangle\otimes|g\rangle)&= \langle \phi|K_l(j)|\psi\rangle, \ \ \textrm{for all }|\phi\rangle,|\psi\rangle\in\mathcal{H}.\label{kkraus}
\end{align}
In the physical notation:
\begin{align}
K_l(j)=\langle \pi_l(j)|U_l|b_l\rangle_{\widetilde{\mathcal{H}}}.\label{15}
\end{align}
\subsection{Description of indirect measurements within Jaynes-Cummings model}
In this section, we specify the description of the indirect measurement of each $l$-th receiver in the frame of the Jaynes-Cummings model \cite{e.t.jaynes} for interaction between a light and a two-level atom. Denote by $|g\rangle$ and $|e\rangle$ -- the ground state and the excited state of a two-level atom, which form an orthonormal basis of $\widetilde{\mathcal{H}}$, and take into account that in (\ref{P}), states
 \begin{align}
|\pi_l(1)\rangle&:=\cos\theta_l|g\rangle+e^{i\xi_l}\sin\theta_l|e\rangle,\nonumber\\
|\pi_l(2)\rangle&:=\sin\theta_l|g\rangle-e^{i\xi_l}\cos\theta_l|e\rangle.\label{14}
\end{align}
admit decompositions. In (\ref{1.75}),
\begin{equation}
\sigma_l=|g\rangle\langle g| \ \ l=1,2,
\end{equation}

According to \cite{e.t.jaynes}, the interaction between a light and a two-level atom is described by the Jaynes-Cummings Hamiltonian on $\mathcal{H}\otimes\widetilde{\mathcal{H}}$:
\begin{equation}
{ H}^{(l)}(t):=H_{0}^{(l)}+H_{int}^{(l)},\label{6}
\end{equation}
where
\begin{equation}
{ H}_{0}^{(l)}:=\hbar\omega_L\left({ a}^\dagger{ a}\otimes\mathbb{ I}_{\widetilde{\mathcal{H}}}\right)+\frac{1}{2}\hbar\omega_0\left(\mathbb{ I}_{\mathcal{H}}\otimes{\sigma_z}\right),\label{free}
\end{equation}
\begin{equation}
H_{int}^{(l)}:=\hbar\Omega_l(t)({ a}\otimes{ \sigma}_++{ a}^\dagger\otimes{\sigma_-}).
\end{equation}
Here, $\omega_L$ is a frequency of the light, $\omega_0$ is a transition frequency of a two-level atom, $\Omega_l(t)$ is a time-dependent interaction parameter, and ${ a}^\dagger$ ($a$) is a creation (annihilation) operator on $\mathcal{H}$ satisfying 
\begin{equation}
{ a}|n\rangle=\sqrt{n}|n-1\rangle, \ { a}^\dagger|n\rangle=\sqrt{n+1}|n+1\rangle \ \ \forall n\in\mathbb{N},
\end{equation}
for every Fock state $|n\rangle$, and ${\sigma}_z$, ${\sigma}_{\pm}$ are the Pauli operators 
\begin{equation}
{\sigma}_z:=|e\rangle\langle e|-|g\rangle\langle g|, \ {\sigma}_+:=|e\rangle\langle g|, \ {\sigma}_-=|g\rangle\langle e|.
\end{equation}
on $\widetilde{\mathcal{H}}$.

In the frame of the Jaynes-Cummings model, in the interaction picture generated by the free Hamiltonian $H_0^{(l)}$, the unitary evolution operator $\widetilde{U}_l$ is the solution of the Schr\"{o}dinger equation:
\begin{equation}
i\hbar\frac{d\widetilde{U}_l}{dt}=\widetilde{H}_{int}^{(l)}(t)\widetilde{U}_l,\label{sch}
\end{equation}
where
\begin{align}
\widetilde{H}_{int}^{(l)}(t):=\hbar\Omega_l(t)\left\{e^{i(\omega_0-\omega_L)t}{ a}\otimes{\sigma}_++e^{-i(\omega_0-\omega_L)t}{ a}^\dagger\otimes{\sigma}_-\right\}\label{7}
\end{align}
is the Jaynes-Cummings interaction Hamiltonian in the interaction picture. If $\omega_L=\omega_0$, then the Hamiltonian (\ref{7}) takes the form
\begin{equation}
\widetilde{H}_{int}^{(l)}(t)=\hbar\Omega_l(t)({ a}\otimes{\sigma_+}+{ a}^\dagger \otimes{\sigma_-}).
\end{equation}
Since
\begin{align}
&\left[\widetilde{H}_{int}^{(l)}(t),\int_0^t\widetilde{H}_{int}^{(l)}(\tau)d\tau\right]=0,
\end{align}
then, as specified in general, for example, in \cite{e.r.loubenets}, the solution of (\ref{sch}) has the form
\begin{equation}
\widetilde{U}_{l}(t):=\exp\left\{-i\widetilde{\Phi}_l(t)({ a}\otimes{\sigma_+}+{ a}^\dagger \otimes{\sigma_-})\right\}\label{12},
\end{equation}
where
\begin{equation}
\widetilde{\Phi}_l(t):=\int_0^t\Omega_l(\tau)d\tau.
\end{equation}
Let us define
\begin{equation}
U_l:=\widetilde{U}_l(T), \ \ \Phi_l:=\widetilde{\Phi}_l(T),
\end{equation}
where $T$ is a time at which the direct measurement on the state $\sigma_l=|g\rangle\langle g|$ on $\widetilde{\mathcal{H}}$ is performed. 

Substituting (\ref{14}) and (\ref{12}) into (\ref{15}), for our case, we derive in Appendix A the following expressions for the Kraus operators (\ref{kkraus}):
\begin{align}
K_l(1)&=\cos\theta_l\cos\left\{\Phi_l\left|a\right|\right\}-ie^{-i\xi_l}\sin\theta_l\sum_{k=0}^{\infty}\frac{(-1)^k}{(2k+1)!}\Phi_l^{2k+1}a\left|a\right|^{2k},\nonumber\\
K_l(2)&=\sin\theta_l\cos\left\{\Phi_l\left|a\right|\right\}+ie^{-i\xi_l}\cos\theta_l\sum_{k=0}^{\infty}\frac{(-1)^k}{(2k+1)!}\Phi_l^{2k+1}a\left|a\right|^{2k},\label{17}
\end{align} 
for each $l=1,2$.

\section{Optimal Success Probability}
In the present section, we specify the above experimental scheme for the case
of binary coherent states $\rho_j:=|\alpha_j\rangle\langle\alpha_j|$, $j=1,2$ where
\begin{equation}
|\alpha_j\rangle=e^{-|\alpha_j|^2/2}\sum_{n=0}^{\infty}\frac{\alpha_j^n}{\sqrt{n!}}|n\rangle.
\end{equation}
Then, the success probability (\ref{17.7}) takes the form:
\begin{equation}
\mathrm{P}_{A|\rightarrow1\rightarrow2}^{success}(|\alpha_1\rangle,|\alpha_2\rangle|q_1,q_2)=\sum_{j=1,2}q_j\left|\left|K_2(j)K_1(j)|\alpha_j\rangle\right|\right|_{\mathcal{H}}^2.\label{21.1}
\end{equation}
In (\ref{21.1}), we derive in Appendix B the following relations:
\begin{align}
K_2(j)K_1(j)|\alpha_j\rangle=\sum_{n=0}^{\infty}F_n(j)|n\rangle,\label{21}
\end{align}
where
\begin{align}
F_n(1)&:=f_n(1)\cos\theta_2\cos\{\Phi_2\sqrt{n}\}-if_{n+1}(1)e^{-i\xi_2}\sin\theta_2\sin\{\Phi_2\sqrt{n+1}\},\nonumber\\
F_n(2)&:=f_n(2)\sin\theta_2\cos\{\Phi_2\sqrt{n}\}+if_{n+1}(2)e^{-i\xi_2}\cos\theta_2\sin\{\Phi_2\sqrt{n+1}\},\label{big_f}
\end{align}
and
\begin{align}
f_n(1)&:=e^{-\frac{|\alpha_1|^2}{2}}\left[\frac{\alpha_1^n}{\sqrt{n!}}\cos\theta_1\cos\{\Phi_1\sqrt{n}\}-i\frac{\alpha_1^{n+1}}{\sqrt{(n+1)!}}e^{-i\xi_1}\sin\theta_1\sin\{\Phi_1\sqrt{n+1}\}\right],\nonumber\\
f_n(2)&:=e^{-\frac{|\alpha_2|^2}{2}}\left[\frac{\alpha_2^n}{\sqrt{n!}}\sin\theta_1\cos\{\Phi_1\sqrt{n}\}+i\frac{\alpha_2^{n+1}}{\sqrt{(n+1)!}}e^{-i\xi_1}\cos\theta_1\sin\{\Phi_1\sqrt{n+1}\}\right].\label{f}
\end{align}

Substituting (\ref{21}) into (\ref{21.1}), we derive
\begin{align}
\mathrm{P}_{A|\rightarrow1\rightarrow2}^{success}(|\alpha_1\rangle,|\alpha_2\rangle|q_1,q_2)=q_1\sum_{n=0}^{\infty}\left|F_n(1)\right|^2+q_2\sum_{n=0}^{\infty}\left|F_n(2)\right|^2\label{24}
\end{align}
and for all $\alpha_1,\alpha_2\in\mathbb{R}$, series in (\ref{24}) converge (see in Appendix C). 
\begin{figure}[ptb]
\includegraphics[width=0.8\linewidth]{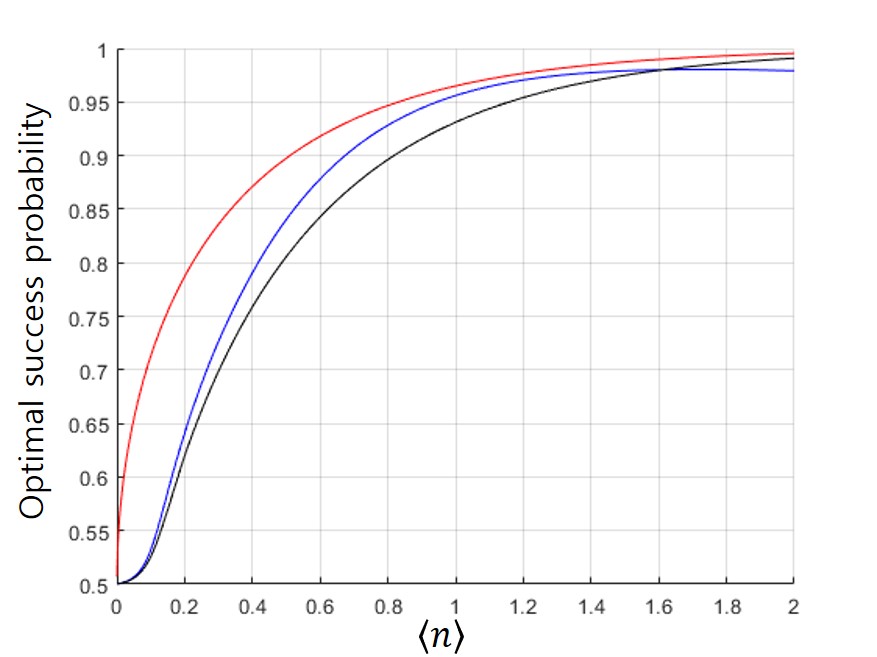}\caption{Optimal success probability of the sequential conclusive state discrimination. Here, $\langle n\rangle$ is a mean photon number. Solid red line, blue line, and black line correspond to the Helstrom bound, our optimal success probability, and the optimal success probability presented in \cite{d.fields}, respectively.}
\end{figure}

Since the success probability (\ref{24}) depends on 
\begin{equation}
{\vec v}=(\Phi_1,\theta_1,\xi_1,\Phi_2,\theta_2,\xi_2)\in\mathbb{R}^6,
\end{equation}
the optimal success probability for the considered protocol is given by the maximum:
\begin{equation}
\mathrm{P}_{A|\rightarrow1\rightarrow2}^{opt.success}(|\alpha_1\rangle,|\alpha_2\rangle|q_1,q_2)=\max_{{\vec v}\in\mathbb{R}^6}\left\{q_1\sum_{n=0}^{\infty}\left|F_n(1)\right|^2+q_2\sum_{n=0}^{\infty}\left|F_n(2)\right|^2\right\}.\label{opt}
\end{equation}

\subsection{Numerical Analysis}
For the numerical analysis of maximum (\ref{opt}), we use Powell's method \cite{m.j.d.powell,j.kiusalaas} realized via MATLAB.\footnote{MATLAB also provides ``fmincon'', which is a command to perform optimization.} 

In Fig.2, we present the results on the numerical  calculation of (\ref{opt}) for the case $q_1=q_2$ and
\begin{equation}
|\alpha_1\rangle=|+\alpha\rangle, \ |\alpha_2\rangle=|-\alpha\rangle, \ \ \alpha>0.
\end{equation}
In this case, the mean photon number is given by
\begin{equation}
\langle n \rangle:=\langle \alpha|a^\dagger a|\alpha \rangle=\langle -\alpha|a^\dagger a|-\alpha \rangle=\alpha^2.
\end{equation}
According to our numerical results presented on Fig.2, 
\begin{itemize}
\item If $\langle n \rangle$ is less than 1.6, then, for our two-sequential state discrimination scheme, the optimal success probability is larger than the one presented in \cite{d.fields}. 
\item Especially, if the $\langle n \rangle$ is almost equal to 1.2, then the optimal success probability nearly approaches the Helstrom bound. 
\end{itemize}

\section{Conclusion}
In the present article, we propose a new experimental scheme for the implementation of the sequential conclusive discrimination between binary coherent states within the Jaynes-Cummings interaction model. We find that if the mean photon number is less than 1.6, then, for our two-sequential state discrimination scheme, the optimal success probability is larger than the one presented in \cite{d.fields}. We also show that, if the mean photon number is almost equal to 1.2, then the optimal success probability nearly approaches the Helstrom bound.

\section*{Acknowledgement}
The study by E.R. Loubenets in Section 2 and 3 of this work is supported by the Russian Science Foundation under the Grant No 19-11-00086 and performed at the Steklov Mathematical Institute of Russian Academy of Sciences. The study by Min Namkung in Sections 2-4 and by E.R. Loubenets in Section 4 is performed at the National Research University Higher School of Economics.

Min Namkung thanks Prof. Younghun Kwon at Hanyang University (ERICA) for his fruitful discussions.

\section*{Appendix A}
In the Appendix, we derive the expression of Kraus operators in (\ref{17}). The unitary operator (\ref{12}) can be expanded in the following infinite series:
\begin{equation}
{U}_l=\sum_{n=0}^{\infty}\frac{1}{n!}\left\{-i{\Phi}_l(a\otimes\sigma_++a^\dagger\otimes\sigma_-)\right\}^n.
\end{equation}
By (\ref{15}), the Kraus operators take the form:
\begin{equation}
K_l(j)=\sum_{n=0}^{\infty}\frac{1}{n!}\langle\pi_l(j)|\left\{-i\Phi_l(a\otimes\sigma_++a^\dagger\otimes\sigma_-)\right\}^n|g\rangle.
\end{equation}
For every $k\in\{0,1,2,3,\cdots\}$, the following relations hold:
\begin{align}
\frac{1}{n!}\left\{-i\Phi_l(a\otimes\sigma_++a^\dagger\otimes\sigma_-)\right\}^n|g\rangle
=\begin{cases} \frac{(-1)^k}{(2k)!}\Phi_l^{2k}|a|^{2k}|g\rangle &\mbox{if } n = 2k \\
-i\frac{(-1)^k}{(2k+1)!}\Phi_l^{2k+1}a|a|^{2k}|e\rangle &\mbox{if } n = 2k+1  \end{cases} \label{30}
\end{align}
Thus, by using (\ref{30}), we derive the following equality:
\begin{equation}
U_l|g\rangle=\cos\left\{\Phi_l|a|\right\}|g\rangle-i\sum_{k=0}^{\infty}\frac{(-1)^k}{(2k+1)!}\Phi_l^{2k+1}a|a|^{2k}|e\rangle.\label{31}
\end{equation}
Substituting (\ref{14}) and (\ref{31}) to (\ref{15}), we derive the Kraus operators (\ref{17}).

\section*{Appendix B}
In the Appendix, we shortly introduce how to derive (\ref{21}). According to (\ref{17}), the following equalities are derived:
\begin{align}
K_l(1)|n\rangle=\begin{cases} \cos\theta_l|0\rangle &\mbox{if } n = 0 \\
\cos\theta_l\cos\left\{\Phi_l\sqrt{n}\right\}|n\rangle-ie^{-i\xi_l}\sin\theta_l\sin\left\{\Phi_l\sqrt{n}\right\}|n-1\rangle &\mbox{if } n \ge 1  \end{cases}\nonumber\\
K_l(2)|n\rangle=\begin{cases} \sin\theta_l|0\rangle &\mbox{if } n = 0 \\
\sin\theta_l\cos\left\{\Phi_l\sqrt{n}\right\}|n\rangle+ie^{-i\xi_l}\cos\theta_l\sin\left\{\Phi_l\sqrt{n}\right\}|n-1\rangle &\mbox{if } n \ge 1  \end{cases}\label{32}
\end{align}
Therefore, by substuting (\ref{32}) to the left hand side of (\ref{21}), we complete the derivation.

\section*{Appendix C}

In the Appendix, we prove that series in (\ref{24}) converge, by using direct comparison test. Firstly, from (\ref{big_f}), it follows that 
\begin{align}
\left|F_n(j)\right|^2\le\left|f_n(j)\right|^2+\left|f_{n+1}(j)\right|^2+2\left|f_n(j)\right|\left|f_{n+1}(j)\right|,
\end{align}
which implies
\begin{align}
\sum_{n=0}^{\infty}\left|F_n(j)\right|^2\le2\sum_{n=0}^{\infty}\left[|f_n(j)|^2+|f_n(j)||f_{n+1}(j)|\right].\label{series}
\end{align}
From (\ref{f}), it follows
\begin{align}
\sum_{n=0}^{\infty}\left|f_n(j)\right|^2&\le e^{-\alpha_j^2}\left\{\sum_{n=0}^{\infty}\frac{\alpha_j^{2n}}{n!}+\sum_{n=0}^{\infty}\frac{\alpha_j^{2n+2}}{(n+1)!}+2\sum_{n=0}^{\infty}\frac{\alpha_j^{2n+1}}{\sqrt{n!(n+1)!}}\right\}\nonumber\\
&\le 2e^{-\alpha_j^2}\left\{\sum_{n=0}^{\infty}\frac{\alpha_j^{2n}}{n!}+\alpha_j\sum_{n=0}^{\infty}\frac{\alpha_j^{2n}}{n!}\right\}\nonumber\\
&=2(1+\left|\alpha_j\right|).
\label{36}
\end{align}
Also, in view of (\ref{f}),
\begin{align}
\sum_{n=0}^{\infty}\left|f_n(j)\right|\left|f_{n+1}(j)\right|\le\sqrt{\sum_{n=0}^{\infty}\left|f_n(j)\right|^2}\sqrt{\sum_{n=0}^{\infty}\left|f_{n+1}(j)\right|^2}\le\sum_{n=0}^{\infty}\left|f_n(j)\right|^2\le2(1+\left|\alpha_j\right|).\label{37}
\end{align}
This proves the convergence of series (\ref{series}) and, correspondingly, the series in (\ref{24}).

\end{document}